\documentclass{article}

\usepackage{geometry}
\usepackage{xspace}
\usepackage[english]{babel}
\usepackage{graphicx}
\usepackage[latin1]{inputenc}
\usepackage{amssymb}

\usepackage{color}

\makeatletter
\@addtoreset{equation}{section}
\makeatother

\author{Jérémie Bourdon and Damien Eveillard}

\title{%
Toll Based Measures for Dynamical Graphs
}

\newtheorem{theorem}{Theorem}

\begin{document}

\newcommand{\fp}{f^{=}}
\newcommand{\fa}{f^{\neq}}
\newcommand{\esp}[2][1]{\hbox{\textbf{E}$\ifnum#1=1\else^{#1}\fi\left [#2\right]$}}
\newcommand{\var}[1]{\hbox{\textbf{Var}$\left [#1\right ]$}}
\newcommand{\Prob}[1]{\hbox{\textrm{Prob}$\left [#1\right ]$}}
\newcommand{\ie}{\textsl{ie.}\ }
\newcommand{\cf}{\textsl{cf.}\ }
\newcommand{\al}{\textsl{\&~al.}\ }
\newcommand{\ex}{\textsl{eg.}\ }
\newcommand{\STARS}{\textsc{Stars}\xspace}
\newcommand{\bsubt}{\textsl{Bacillus~Subtilis}\xspace}

\def\pn{\par\noindent}
\def\Ind{\hbox{{\large $1\kern-3pt$I}}}

\def \W {{\cal L}}
\def\pn{\par\noindent}
\def\II{{\cal I}}
\def\IJ{{\cal J}}
\def\LL{{\cal L}}
\def\OA{\mathbb{A}}
\def\OB{{\bf B}}
\def\OC{{\bf C}}
\def\OD{{\bf D}}
\def\OG{{\bf G}}
\def\OH{{\bf H}}
\def\OK{{\bf K}}
\def\OL{{\bf L}}
\def\OM{{\bf M}}
\def\OP{\mathbb{P}}
\def\ON{\mathbb{N}}
\def\OR{{\bf R}}
\def\OS{{\bf S}}
\def\OT{\mathbb{T}}
\def\OX{{\bf X}}
\def\G {{\bf G}}
\def \E {{\cal E}}
\def \W {{\cal W}}
\def\SG{{\mathbb G}}
\def\GM{{\mathbb M}}
\def\GG{{\mathbb G}}
\def\MG{{\mathfrak G}}
\def\MH{{\mathfrak H}}
\def\MN{{\mathfrak N}}
\def\MP{{\mathfrak P}}
\def\Sp{\hbox{Sp}\!}
\def\induit{{\widetilde{\MG}}}
\def\carac{\hbox{{\large $1\kern-3pt$I}}}
\def\Ind{\hbox{{\large $1\kern-3pt$I}}}
\def\esp#1{\mathbb{E}\left [#1\right ]}
\def\Prob#1{{\hbox{\textrm{Prob}}}\left\lbrace #1 \right \rbrace}
\def\Es {{\mathbb E}}
\def\var{{\mathbb V}}
\def\overbar{\overline}
\def\ds{\displaystyle}
\def\R{\mathbb R}
\def\LD{{\cal D}}

\def\sign{\hbox{sgn\ }}

\def\pattr{\hbox{\rm PaTr\ }}
\def\shuffle{\amalg}
\def\pattern{\alpha}
\def\patternII{w}
\def\infil{\uparrow}
\def\MB{{\mathfrak B}}
\def\MG{{\mathfrak G}}
\def\MH{{\mathfrak H}}
\def\RG{\hbox{\textbf{G}}}
\def\RP{\hbox{\textbf{P}}}
\def\RN{\hbox{\textbf{N}}}
\def\RL{\hbox{\textbf{L}}}
\def\II{{\cal I}}
\def\Tr{\hbox{Tr\ }}
\def\A{\Sigma}
\def \cal {\mathcal}
\def\L{{\cal L}}
\def\M{{\cal M}}
\def\B{{\cal B}}
\def\C{{\cal C}}
\def\N {{\cal N}}
\def\w {{\cal W}}
\def\CD {{\cal D}}
\def \E {\mathcal{E}}
\def\CP {{\cal P}}
\def\LD{{\cal D}}
\def\LF{{\cal F}}
\def\LG{{\mathbf G}}
\def\LI{{\mathbf I}}
\def\LL{{\mathbf L}}
\def\LN{{\mathbf N}}
\def\LO{{\cal O}}
\def\LP{{\mathbf P}}
\def\LQ{{\cal Q}}
\def\LR{{\mathbf R}}
\def\LS{{\cal S}}
\def\I{{\cal I}}
\def \Var {\bf   Var \, }
\def \ds {\displaystyle}
\def\x {\mathbb{X}}

\def\alphabet{\mathcal{A}}

\def\varemptyset{\oslash}
\def\card{\hbox{\rm card}}
\def \pn {\par \noindent}
\def\CQFD{
\vrule height 3pt depth 3pt width 1pt\raise 3pt\hbox{\vrule height 0pt
depth 1pt width 4pt}\kern -4pt\raise -3pt\hbox{\vrule height 1pt depth
0pt width 4pt}\vrule height 3pt depth 3pt width 1pt
}
\def\restr#1{\vrule height 8pt depth 5pt width 
0.4pt\raise-5pt\hbox{$\scriptstyle #1$}}

\maketitle


\begin{abstract}
	Biological networks are one of the most studied object in computational biology. Several methods have been developed for studying qualitative properties of biological networks. Last decade had seen the improvement of molecular techniques that make quantitative analyses reachable. One of the major biological modelling goals is therefore to deal with the quantitative aspect of biological graphs.	
	We propose a probabilistic model that suits with this quantitative aspects. Our model combines graph with several dynamical sources. It emphazises various asymptotic statistical properties that might be useful for giving biological insights.

\pn\textbf{Keywords:} average-case analysis; Biological networks; Markov chains; Dynamical sources.
\end{abstract}

  \section{Introduction}
  \label{sec:introduction}
Biology is concerned with the study of  biological functions that change chemical components into another. The biological system that provides such  transformations produces both biological matter and energy. Genetic and biochemical investigations during the last decade have changed our understanding of biological transformation processes. Today, molecular techniques allow the description of complex networks of interacting macromolecules that are responsible of theses phenomenological transformations. Understanding biological molecular network is therefore one of the major goal of the modern biology. Systems biology uses computational approaches in this purpose (see \cite{Szallasi} for overview). In this context, our aim is studying the dynamic of molecular interaction networks.

Until now, precise and quantitative informations on reaction mechanisms have been roughly available for most networks of interest. Therefore, based on different formalisms, various studies have focused on the qualitative modelling of dynamical networks \cite{Hidde02}.  Furthermore, because biological knowledge is incomplete, qualitative modelling has appeared as an appropriate framework for reasoning on the biological network and refining the graph whenever additional biological knowledge becomes available. By now novel methods in biological data acquisition introduce the ability of high throughput measurements that efficiently map out a network of interacting macromolecules. It thus gives various quantitative insights of biological systems. One of the major biological modelling goals is therefore to deal with the quantitative aspect of  dynamical networks with a special emphasis on the quantitative reasoning that naturally follows the qualitative reasoning. We propose here a novel probabilistic model that suits this quantitative aim. We consider the cost of each macromolecular interaction for emphasizing the major quantitative pathways among others qualitative. 

Formerly, we study the typical behavior of the cost on a pathway that pass through the biological graph. These costs are sums of elementary (fixed) toll costs on the edges taken by paths. This measure on a graph $G$ with $r$ vertices can be exactly described by a $r\times r$ real matrix whose element $(i,j)$ is the toll cost of the edge $(i,j)$. Later, this matrix is denoted by $C(G)$. We aim at studying random costs variables, it is thus necessary to ensure that they are not  degenerated. A toll cost matrix $C(G)$ is said to be non degenerated if for all strictly positive integer $n$, there exists at least two paths of length $n$ in graph $G$ whose measures differ.

When a path is taken at random over some probabilistic model, the total cost itself is a real random variable.
A simple probabilistic framework for generating random paths on a graph is to weight properly the edges of the graph by some fixed probabilities. This model is a classical Markov chain model. In their book~\cite{meyn}, Meyn \textsl{et al.} provides a general review of classical results on paths in Markov chains.

\medskip

Here, we define a more general probabilistic framework that in some sense extends Markov chains.
The edges of the graph are weighted by some dynamical sources, a general model presented in the context of pattern matching in~\cite{bovalatin,generalized,CMV,Va1}. 

Dynamical sources describe non-Markovian processes,  
where the dependency on  past history is unbounded,   
and as such, they reach a high level of generality. 
    A probabilistic dynamical source  is defined by two objects: $(i)$  symbolic 
mechanism and $(ii)$ density.
The mechanism, related to symbolic dynamics,     associates an infinite word $M(x)$ 
to a real number $x\in [0, 1]$,
 and  generalizes numeration systems. 
  Once the mechanism   has been fixed,  the  density  on  the 
 $[0, 1]$ interval  can  vary. This induces different  
probabilistic behaviours for sources of words. Later, we establish a correspondence between paths and words over an appropriate alphabet.

Therefore, a dynamical graph is defined by the combination of a graph and several dynamical sources. Like this, one dynamical source is associated with each vertex of the original graph.

In this context, it is proved that the total cost of a path in the dynamical graph follows
 asymptotically a gaussian law.

\smallskip
\pn {\bf Definition 1.} [Asymptotic gaussian law.] {\sl Consider a cost
$R$ defined on  a set $\mathcal{R}$ and its restriction $R_n$ to  the subset $\mathcal{R}_n$ of size $n$.  The
cost $R$ asymptotically follows a gaussian law as $n\rightarrow +\infty$ if there exist three
sequences $a_n, b_n, r_n$, with $r_n\to 0$, for which
\vspace{-0.2cm} $$ \Pr  \left [(u, v) \in \mathcal{R}_n~\bigm|~
\frac {R _n(u, v) - a_n}{    \sqrt{  b_n} }\le y
\right]  =
\frac {1} {\sqrt {2\pi} }
\int_{- \infty}^y e^{-t^2/2} \, dt + O(r[R_n])\, .
$$
\vskip -0.2 truecm
\pn The sequence $r_n$ defines the speed of convergence, denoted also
by $r[R_n]$. The expectation $\Es[R_n]$ and the variance $\var[R_n]$
satisfy $\Es[R_n] \sim a_n, \quad \var[R_n] \sim b_n$.  The
 triple  $(\Es[R_n], \var[R_n] , r[R_n])$ is a characteristic triple for the gaussian
law of $R$.}

\smallskip
\pn {\bf Main Theorem.} {\sl 
Let  ${\cal G}=(G,\mathcal D)$ be  a nice\footnote{The word ``nice'' is defined in Def. 4, 
Section 4.1, the word ``primitive'' in Section 4.1}  dynamical graph and $C(G)$ be a non degenerated toll cost matrix on graph $G$.

\pn 
The total cost of a path of length $n$ in the dynamical graph ${\cal G}$, denoted by   $C_n({\cal G})$, follows an asymptotic gaussian law as $n\rightarrow+\infty$ with a characteristic triple given  by   \quad $ r[C_n(\mathcal{G})]  = O(1/\sqrt n)$
  
 \vspace {0.2 truecm}
\centerline{ $
\Es[ C_n(\mathcal{G})] =   \gamma_{\cal G} \cdot n  +  \gamma'_{\cal G} + O(\mu_{\cal G}^n)
\quad\hbox{ and }
\var[ C_n(\mathcal{G})] =    \nu_{\cal G} \cdot n  +  \nu'_{\cal G}  + O(\mu_{\cal G}^n)$}
 \vspace {0.1 truecm}
\pn The constants  $\gamma_{\cal G}$ and $\nu_{\cal G}$ are expressible  with  the  pression  $\Lambda(t)$ of the operator  $\mathbb{T}(e^t)$ defined  in (\ref{defT}), namely 
$
 \gamma_{\cal G} = \Lambda'(0),
\ 
 \nu_{\cal G}  = \Lambda''(0),   
$
 while $\mu_{\cal G}<1$ is any real number strictly larger than the subdominant eigenvalue of $\mathbb{T}$.
}

\medskip
Such a definition leads to formulate various tools.

  \section{Various tools}
  \label{sec:dyngraphs}
  
  We propose a probabilistic model for paths that compose a randomized graph. We relate the study of basic parameters of these paths with formal generating functions.
  It defines the dynamical sources model and provides major properties for their related generating operators.
  
  \subsection{Probabilistic model and generating functions}
Let $G=(V,E)$ an oriented, strongly connected graph with $r$ vertices $V=\{1,\dots,r\}$. A path of length $n$ in the graph in a $n$-uple $\mathcal{W}=w_1\rightarrow w_2\rightarrow \dots \rightarrow w_n$, where $w_i\in V$ for all $i\in\{1,\dots,n\}$ and $(w_i,w_{i+1})\in E$ is an edge for all $i\in\{1,\dots,n-1\}$. We denote by $\mathcal{W}_n(G)$ the subset of $V^n$ composed by all possible paths of length $n$ in graph $G$ and by $\mathcal{W}_\star(G)$ the subset of $V^\star$ composed by all possible paths of any length in graph $G$.

We focus on graphs that are strongly connected (i.e., for all pairs of vertices $(i,j)$, there exists a path that links $i$ to $j$. This technical request is crucial since it ensures that the transition matrix of the graph is primitive and $\mathcal{W}_n(G)$ as well.

Random paths are drawn according to an induced probability on $\mathcal{W}_\star(G)$. In the sequel, we denote by $p_W$ the probability that a random (infinite) path begins with the (finite) path $W$ (of length $n$).

Now, a measure on these paths is a function $c:\mathcal{W}_\star(G)\rightarrow\mathbb{R}$ whose restriction $C_n$ to 
$\mathcal{W}_n(\mathcal{G})$ is a real random variable. Our aim is to analyze its probabilistic behavior as $n\rightarrow +\infty$, when $c$ is a measure that can be described by toll cost values on edges (i.e., each edge contributes by a fixed value to the total measure).
Our main tool is the moment generating function of $C_n$, 
defined as
\begin{equation}
\label{mogefu} 
\esp{\exp (tC_n)} : = \sum_{ W \in \mathcal{W}_n(G)}  p_W \cdot \exp [{t c(W)}], 
\end{equation}
 and  the major issue is showing that it behaves  as a ``quasi-power''.  Then, Hwang's quasi-power theorem~\cite{hwang} is used to conclude to an asymptotic Gaussian law.

\medskip
\begin{theorem}[Hwang]
{\sl Let $Y_n$ be a sequence of  variables whose moment generating functions satisfies when $n\rightarrow+\infty$
$
\esp{\exp(t Y_n)}= [\exp (nU(t)+V(t))] \cdot [1+O(W_n)],\ \ 
W_n\rightarrow\infty,
$
 with a  uniform error term   on the complex closed disk $D(t_0):=\{|t|\leq t_0\}$, $t_0>0$.
Suppose that  $U(t)$ and $V(t)$ are analytic in  $D(t_0)$ and $U(t)$ satisfies $U''(0)\neq 0$.
Then,  $Y_n$ follows an asymptotic gaussian law, with a characteristic triple given by 
\vspace{-0.2 cm}
$$
\esp{Y_n} = U'(0)\cdot n  +  V'(0)+ O(W_n),
\quad
\var{[Y_n]} =U''(0) \cdot n  + V''(0) + O(W_n), 
$$
\centerline{$\quad r[Y_n] = O\left( \max(1/{\sqrt{n}},W_n)\right).$}  }
\end{theorem}

\subsection{Bivariate generating functions}
The  so--called   probability  generating function $F_c(z, u)$ relative to parameter $c$ is   defined as
\begin{equation}
\label{generatingfunction}
F_c(z,u)=\sum_{W\in \mathcal{W}_\star(G)} p_W \cdot u^{c(W)}\cdot  z^{|W|},
\end{equation}
where $|W|$ denotes the length of the path $W$, the variables $z$ and $u$ respectively mark  the length of the path and  the parameter $c(W)$. Remark that the moment generating function of parameter $C_n$ is closely related to $F_c(z, u)$ via the relation 
\begin{equation}
 \esp{\exp(t C_n)} = [z^n] F_c(z, e^t)
 \end{equation}
where the notation $[z^n] G(z)$ denotes the coefficient of $z^n$ in $G(z)$. 
We now express the generating function $F_c(z,u)$ in a form that simplify the coefficient extraction.

\subsection{Transition matrix and generating functions}
The transition matrix $T:=(T_{i,j})$ of the graph $G$ (with $r$ vertices) is the $r\times r$ matrix which element  of index $(i,j)$ equals $1$ iff $(i,j)$ is an edge of $G$ (and 0 otherwise). Thus, $T^n$ is a $\{0,1\}$ matrix that indicates if there exists at most one path in the graph from one state to another.

Focussing on the paths, it is convenient to deal with a marked version of the transition matrix $\mathcal{T}:=(\mathcal{T}_{i,j})$ such that: $\mathcal{T}_{i,j}=e_{i,j}$ if $T_{i,j}=1$ and $\mathcal{T}_{i,j}=0$ otherwise. In the sequel, $\Sigma=\Sigma(G):=\{e_{i,j}| (i,j)\in E\}$ refers to the set of all possible edges, and $\Sigma_i:=\{e_{i,j}, (i,j)\in E\}$ refers to the edges that leave vertex $i$. Notice that a path in $G$ corresponds to a string over $\Sigma^\star$.
The matrix $\mathcal{T}$ plays a fundamental role.
Indeed, the component $(i, j)$ of the matrix ${\cal T}^n$ is the set formed by all the paths of length $n$ which allow to reach state $j$ from state $i$. In addition, the component $(i, j)$ of the matrix  ${\cal T}^\star$
is the set formed by all possible paths that allow to reach state $j$ from state $i$ using an arbitrary number of steps. Finally, the set $\mathcal{W}_n(G)$ of all possible paths of length $n$ is expressed formally by
\begin{equation}
\label{eq:tpuissn}
\mathcal{W}_n(G)=  \mathbf{1} \cdot {\cal T}^n \cdot {}^t\mathbf{1}, \quad \mathcal{W}_\star(G) =  \mathbf{1} \cdot {\cal T}^\star \cdot {}^t\mathbf{1},  \quad 
\end{equation}
where $\mathbf{1}:= (1,\dots, 1)$ is the $r$ length vector of $1$.

\medskip
We  now present the probabilistic model for path generation. This model  is based  on dynamical  systems.  
 Probabilities of passing through an edge are ``generated'' by operators, and the main generating function can be generated itself  by  operators. Furthermore, unions and Cartesian products of sets translate
into sums and compositions of the associated operators.  This allows to define a matrix generating operator  related to a graph and its associated probabilistic model.

\subsection{Dynamical sources}
We first recall the definition of a dynamical system (of the interval) (See~\cite{Va1,CMV} for details). 
\smallskip
\pn {\textbf{Definition 2.}}
{\sl A {\em dynamical system} $\mathcal{D}=({\cal I}, {\cal S})$ is defined by four 
elements:

 $(a)$ {a finite alphabet $\alphabet$,}

 $(b)$ {a topological partition of ${\cal I}:=]0,1[$
with disjoint open intervals ${\cal I}_m,m\in\alphabet$,}

$(c)$ {an encoding mapping $\sigma$ which is constant and equal 
to $m$ on
each ${\cal I}_m$,}
 
$(d)$ {a shift mapping $S$ whose restriction to ${\cal I}_m$ is 
a {bijection} of class ${\cal C}^2$  from ${\cal I}_m$ to ${\cal J}_m:=S(\mathcal{I}_m)$. The    local inverse    of 
 $S|_{{\cal I}_m}$  is denoted by $h_m$.}
 }

 \smallskip
 \pn  Such  a dynamical system can be viewed as a ``dynamical source", since, 
 on an input $x$ of ${\cal I}$, it outputs the word 
 $M(x)$ formed by
the sequence of symbols $\sigma S^j(x)$, i.e., 
$
M(x):=(\sigma x,\sigma S x,\sigma S^2 x,\dots).
$

\pn The  branches of $S^k$, and also its inverse branches,  are then indexed 
by ${\alphabet}^k$, and, for any  $w= m_1\dots m_k \in {\alphabet}^k$, the mapping
 $h_w:=h_{m_1}\circ h_{m_2}\circ \cdots \circ 
h_{m_k}$ is a  $\mathcal{C}^2$ {bijection} from  $\mathcal{J}_w$ onto  $\mathcal{I}_w$. It is possible that the word $w$ cannot be produced by the source: this means that ${\cal J}_w$ is empty, and the inverse branch $h_w$ does not exist.  All the  words that begin with the same prefix $w$ correspond 
to real numbers $x$ that belong to the same  interval 
${\cal I}_w$.  

\smallskip
\pn Such sources may possess a high degree of correlations, due to the {\sl geometry} of the branches  [i.e., the respective positions of intervals ${\cal I}_m$ and ${\cal J}_\ell := S({\cal I}_\ell)$] and also to the {\sl shape} of branches [See \cite{CMV} for more details]. For instance, classical sources correspond to dynamical systems with affine branches, for which the derivatives are constant.
In other words,  the probability of emitting a symbol $m$ is closely related  to the shape of branches.

\medskip
\subsection{Probabilities and generating operators}
When the interval ${\cal I}$ is endowed with some density $g$, this induces a probabilistic model 
on  $\alphabet^{\mathbb N}$, and 
 the probability $p_w$ that a word begins with prefix $w$
is the 
measure of the interval ${\cal I}_w$.  Such a probability $p_w$  
is  easily generated  by  an  operator $\mathbf{G}_{[w]}$,  defined as
\vspace{-0.2 truecm}
\begin{equation} \label{pw}
 {\bf G}_{[w]}[f](x)  =  |h'_w(x)|   \ f\circ h_w (x)\Ind_{\mathcal{J}_w}(x), 
  \end{equation}
\vspace{-0.4 truecm}
\begin{equation}
\label{eq:intpw}
 \hbox{since  one has} \quad p_w= \int_{\mathcal{I}_w} g(x) dx= \int_{\mathcal{J}_w} |h'_{w}(x)|  g\circ 
h_{w}(x) dx =  \int_{0}^1 {\bf G}_{[w]}[g](x)  dx.
\end{equation}
 The operator $\mathbf{G}_{[w]}$ is called the generating operator of the prefix $w$.  The  generating operator  ${\bf L}$
relative to a collection  ${\cal L}$ of words  is defined as    the  sum of all the generating operators relative to the 
words of ${\cal L}$, namely  $\mathbf {L}  := \sum_{w \in {\cal L}}   \mathbf{G}_{[w]}$,  and   the   generating operator ${\bf G}$ of the alphabet $\alphabet$ 
\vspace{-0.2 truecm}
\begin {equation} \label {G}
  \qquad  { \bf G}:=\sum_{m \in \alphabet} {\bf G}_{[m]}.
  \end{equation}
  \vspace{-0.4 truecm} \pn
plays a fundamental role  here, 
since it is the density transformer of the dynamical system;   it describes  the evolution of densities on ${\cal I}$ under iterations of $S$:  if $X$ is a random variable with density $g$,  then $SX$ has density ${\bf G}[g]$.   
  \pn 
 For  two prefixes $w, w'$, the relation $p_{w.w'}= p_w p_{w'}$  is no longer true
when the source has  some memory, and is replaced by the following  composition property 
\vspace{-0.3 truecm}
\begin{equation} \label {Gww'}
\mathbf{G}_{[w.w']}=\mathbf{G}_{[w']}\circ\mathbf{G}_{[w]}, 
 \end{equation}
 \vspace{-0.4 truecm}\pn 
so that  unions and Cartesian products 
of collections of words  translate
into sums and compositions of the associated  generating operators.    Remark just that, due to (\ref{Gww'}),    the generating operator  of ${\cal L}\times {\cal M}$ is $ \mathbf {M} \circ \mathbf {L}$. 

\section{Dynamical graph}
We construct a probabilistic model for paths by combining both information of the graph and information of several dynamical systems. This provides the so-called \textsl{dynamical graph}.

A \textsl{dynamical graph} $\mathcal{G}$ is defined by a pair $(G,\mathcal{D})$ where $G=(V,E)$ is an oriented strongly connected graph with $r$ vertices ($V=\{1,\dots,r\}$) and $\mathcal{D}=(\mathcal{D}_1,\dots,\mathcal{D}_r)$ is a sequence of $r$ dynamical sources. The dynamical source $\mathcal{D}_i$ corresponds to the vertex $i$. It is defined by the alphabet $\Sigma_i:=\{e_{i,j}, (i,j)\in E\}$ and describes the probability of leaving the vertex $i$. Indeed, previously we suggested that the dynamics of $\mathcal{D}_i$ are properly represented by generating operators $\mathbf{G}_{i,[e]}$, $e$ is any symbol/edge of $\Sigma_i$. Thus, if at a time, we are in vertex $i$ with some density $g$ that corresponds to all the previous edges traversed by the path before $i$, the probability of taking the edge $(i,j)$ is given by
$$
p_{j|i,g} = \int_0^1 \mathbf{G}_{i,[e_{i,j}]} [g](t) dt.
$$

\pn\textbf{Remark. } When all the dynamical systems $\mathcal{D}_i$ are simple Bernoulli sources, the dynamical graph is a Markov chain (of order $1$). In this case, if there exists an edge from vertex $i$ to vertex $j$, the probability of taking the edge $(i,j)$ is the conditional probability $p_{j|i}$.
Thus, dynamical graphs extend the Markov process model by introducing a high level of correlations between the different edges taken by the path.

Here, we present two distinct tools dedicated to dynamical graphs. The first is a matrix generating operator. The second is the (flat) generating operator of a (mixed) dynamical source closely related to the dynamical graph. These two objects are conjugated and share the same spectrum.

\subsection{Matrix generating operator}

We transform the transition matrix of a graph into a matrix generating operator that combines both information from the graph and the different dynamical sources of vertices. We  associate to
   each element ${\cal T}_{j, i}$ of the marked matrix $\mathcal{T}$, its  generating operator $\mathbb{T}_{i,j}$
 \begin{equation}\label {OT}
 \mathbb{T} _{i,j} :=  \mathbf{G}_{i,[e_{i,j}]}
 \end{equation}
$\mathbb{T}$ is  a matrix generating operator which is related to  $^t {\cal T}$,  due to (\ref{Gww'}).

In order to represent properly the generating function of the parameter $c$, when $c$ is a toll based measure, we  introduce some ``perturbations'' in the matrix operator. The perturbed matrix operator $\mathbb{T}(u)$ is a matrix operator whose elements are
\begin{equation}
\label{defT}
\mathbb{T}_{i,j}(u):=u^{c_{i,j}} \mathbb{T}_{i,j},
\end{equation}
where $c_{i,j}$ is the toll cost of edge $(i,j)$.
Obviously, $\mathbb{T}(1)$ corresponds to the unperturbed version of the matrix operator.
Thanks to $(\ref{eq:tpuissn})$ and $(\ref{eq:intpw})$, we express $F_c(z,u)$ by means of the perturbed matrix operator $\mathbb{T}(u)$,
$$
F_c(z,u)=\int_0^1 \mathbf{1} \cdot  (I-z \mathbb{T}(u))^{-1}\cdot {}^t\mathbf{1} [g](x) dx,
$$
where $g$ is an initial given density.
The coefficient of $z^n$ in $F_c(z,u)$ corresponds to the $n$-th power of the matrix operator, consequently,
the moment generating function of the cost $C_n$ satisfies
\begin{equation}
\label{momentn}
\esp{\exp(tC_n)}=\int_0^1 \mathbf{1} \cdot  \mathbb{T}(e^t)^n \cdot {}^t\mathbf{1} [g](x) dx.
\end{equation}

\subsection{Mixed source}

We  build a source ${\cal S}_{\cal G}$ that combines both a transition matrix  $T$ of a dynamical graph ${\cal G}$, and the original sources ${\cal D}_1,\dots,{\cal D}_r$. The set of vertices of the underlying graph $G$ is $V:=\{1,\dots, r\}$ and the transition matrix $T$ of $G$ has order $r$. For all $i$, the source ${\cal D}_i$  is defined by an interval $\II^{(i)}=[0,1]$, an alphabet $\Sigma_i$, a topological partition $(\II^{(i)}_m)_{m\in\Sigma_i}$ and a shift $S^{(i)}$ which local inverse $h^{(i)}_m:=(S^{(i)}\restr{\II^{(i)}_m})^{-1}$  maps $\IJ^{(i)}_m:=]c^{(i)}_m, d^{(i)}_m[$ on $\II^{(i)}_m:=]a^{(i)}_m, b^{(i)}_m[$.

\pn The source ${\cal S}_{{\cal G}}$ is defined with the interval $\II_{\mathcal{G}}=[0,r]$, the alphabet $\Sigma:=\bigcup_{i=1}^r \Sigma_i$, a topological partition $(\II_{m})_{m\in\Sigma}$ and a shift function $S_{\mathcal{G}}$ that maps $\II_{\mathcal{G}}$ on $\II_{\mathcal{G}}$. Each  local inverse $h_m$ maps $\IJ_m$ on $\II_m$.  More precisely, if $m=e_{i,j}\in\Sigma_i$, 
$$
\II_{m}=\II^{(i)}_m + i -1 := ]a^{(i)}_m+i-1, b^{(i)}_m+i-1[,
\quad
\IJ_m=\IJ^{(i)}_m + j-1 := ]c_m+j-1, d_m+j-1[,
$$
$\hbox{ and }
h_m(x)=h_m(x-j-1)+i-1.
$
 The density transformer $\MG$ of the source ${\cal S}_{\cal G}$   defined,  as in (\ref{G}),  by
\vspace{-0.2 truecm}
\begin{equation} \label {MG}
\MG[f](x):=\sum_{m\in\Sigma} |h_m'(x)|\cdot  f\circ h_m(x)\cdot \Ind_{\mathcal{J}_{m}}(x),
\end{equation}
\vspace{-0.4 truecm}\pn 
 is conjugated to  the matrix operator $\OT$  defined in  (\ref{OT})  via  a mapping $\Psi$ [namely
$ 
\MG=\Psi ^{-1}\circ \OT \circ \Psi
$]
 which   associates to $g$  (defined on $\II$)  the vector $^t [ g_1,  \ldots,  g_r  ]$ where each $g_i$ is defined  on $\II^{(i)} $ by $g_i(x):=g\restr{[i-1,1]}(x+i)$. 

\section {Probabilistic behavior of toll based measures}
For studying the parameter of interest, the major issue is to
 prove that both graph and dynamical systems possess good properties. It is the same for the source $\mathcal{S}_{\mathcal{G}}$ associated to the dynamical graph $\mathcal{G}=(G,\mathcal{D})$.
As mentioned in~\cite{bovalatin}, we consider dynamical automatons and extends the results to dynamical graphs.

\subsection{Nice sources and convenient sources.}
Under  general hypotheses,  and on a convenient functional space, the density transformer admits $\lambda = 1$ as an  eigenvalue of 
largest modulus. Nevertheless, this is not a  unique dominant eigenvalue isolated from the remainder of the spectrum. 

\smallskip
\pn  {\bf Definition 3.} 
{\sl A dynamical source is   said to be decomposable if   
the  density transformer  ${\bf G}$  [defined in (\ref{G})]   has  a unique dominant eigenvalue (equal to 1) 
separated from the remainder of the spectrum by a spectral gap,  i.e., $\rho := \sup \{  |\lambda|\ \ ; \quad\lambda \in {\rm Sp}\,  {\bf G} , \lambda \not = 1 \} <1$, when acting on a convenient Banach  space ${\cal F}$.}

\smallskip
\pn {\bf Remarks.}  The terminology  considers the dominant
eigenfunction 
 $\varphi$ which is an invariant function for ${\bf G}$.
Under the normalization condition $\int_0^1\varphi(t)dt=1$, this last  
object is unique too, and it is also  the (unique) stationary density.   Due   
to the existence of the spectral gap,    the operator ${\LG}$ decomposes into two parts, namely  $ \LG = \lambda \LP + \LN$, where 
$\LP$ is the projection of $\LG$ onto the dominant 
eigenspace generated by $\varphi$, and $\LN$,  relative to the remainder of the spectrum, 
has a spectral radius  equal to $\rho$, which is strictly  less than 1.   The operator   $\LN$ describes
the correlations of the source.  A  decomposable dynamical source is   ergodic and 
mixing with  an exponential rate equal to  $\rho$.  

\medskip
\pn  Most of the classical sources --memoryless sources, 
or  primitive Markov chains-- are easily proven to be decomposable.  
  We present sufficient conditions under which a  general dynamical source  is proven to be decomposable, together with all its associated mixed sources
${\cal S}_{\cal G}$ [the proofs are   omitted here].

\medskip
\pn {\bf Definition 4.} 
{\sl A dynamical source (on a finite alphabet) is said to be ``nice''   if it satisfies the two conditions

\pn $(i)$  {\rm [Expansiveness]}  There  exist two constants $C, D$ with $ D>1$ for which one has, for any $m \in {\Sigma}$, for any $x \in {\cal I}_m$, $D< |S'(x)]< C$.  

\pn $(ii)$ {\rm [Topologically mixing]} For any  pair of two nonempty open sets $(V, W)$,   there exists $n_{0} \ge 1$
such that  $S^{-n} V \cap W \not = \emptyset $ for all $n \ge
n_{0}$. }

\medskip
\pn {\bf Proposition 1.} {\sl A nice  dynamical system is decomposable, with respect to the   space 
$BV({\cal I})$  of functions with bounded variation, endowed with the norm \ 
$ ||f|| :=  \sup |f| + V(f)$ [Here,    $V(f)$ is the total 
variation  of $f$ on ${\cal I}$].}

\smallskip
\pn We consider now the  mixed source ${\cal S}_{\cal G}$.
Note here   that a  transition matrix $\mathcal{T}$ is primitive if there exists a power of the matrix $\mathcal{T}$ whose coefficients are never the empty language. A strongly connected  graph  produces a matrix ${\cal T}$, primitive if and only if  the gcds of the lengths of its cycles equals 1.  If 
it is not primitive, the gcd  $d$ of its cycle lengths is  called the period, and  ${\cal T}^d$ is primitive.

\medskip
\pn {\bf Proposition 2.} {\sl Let $\mathcal{G}=(G,\mathcal{D})$ be a dynamical graph. If ${\cal D}$ is a sequence of nice dynamical sources and the transition matrix of $G$ is primitive, the mixed source ${\cal S}_{\cal G}$ is nice too.
}

\subsection{Proof of the main theorem.}
The main theorem is then a consequence of all the previous facts. Indeed, thanks to Propositions 2 and 3
the density transformer $\mathfrak{G}$ of the mixed source $\mathcal{S}_{\mathcal{G}}$ has dominant spectral properties, and by conjugation and perturbation theory, this transmits to the quasi-inverses of marked operator $\mathbb{T}(u)$, when $u$ remains in a neighbourhood of 1. Finally, $\mathbb{T}(u)$ admits the decomposition
$$
\mathbb{T}(u)^n = \lambda(u)^n \mathbb{P} + \mathbb{N}(u),
$$
in a complex neighbourhood of $u=1$. Then, with $(\ref{momentn})$,
the moment generating function of the total cost $C_n$ behaves as an approximate $n$-th power. We conclude by using Hwang's quasi-power Theorem~\cite{hwang} (See $2.1$).

\section{Examples of toll based measure schemes}
    \label{sec:scheme}

We present here several measure schemes. They all consist in a sum of elementary (fixed) toll costs on the edges taken by the path. Thus, a measure on a graph $G$ with $r$ vertices can be exactly described by a $r\times r$ real matrix whose element $(i,j)$ is the toll cost of the edge $(i,j)$. 
    
    \subsection{Counters on edges}
Here, we want to have access to statistics on the number of times a given edge is taken.
It is surely the simplest measuring scheme. Nevertheless, it is quite important since it is the basis of all the next measure schemes.
It consists in marking a single transition $(i,j)$ by a cost equal to $1$. Thus the toll cost matrix is a matrix of zeros except at the position $(i,j)$.
        
    \subsection{Counters on vertices}
In order to count the number of times we reach a given vertex, it is sufficient to consider all the edges that point to this vertex. More formerly, the number of times a given vertex $k$ is reached is obtained by using a toll cost matrix such that $c_{i,k}=1$ if $(i,k)\in E$ and equals 0 otherwise.

From an applicative point of view, it also proves useful to have some real toll costs for the edges that point to a given vertex. This allows to reflect for instance the stoichiometry of a reaction (represented by an edge) between products (represented by the vertices).

    \subsection{Counters on pathways}
The last measure scheme consists in counting particular pathways (here, a pathway is a possible (finite) succession of several edges in the graph that does not contain any cycle). 
This kind of statistic is useful to compare different pathways that link the same vertices in a graph.

This measure is done by duplicating all the vertices along the pathway. This new graph is called $G'$. The copy is used to count the pathway of interest (by counting its last transition) which is removed from the original graph (by removing its first transition). Finally, a copy of appropriate edges (all the edges that leave a vertex of the path) ensures that two graphs $G$ and $G'$ share the same sets of paths (i.e., $\mathcal{W}_n(G')=\mathcal{W}_n(G)$ for all $n$ and $\mathcal{W}_\star(G')=\mathcal{W}_\star(G)$).

More formerly, let $P=i_0\rightarrow i_1\rightarrow \cdots \rightarrow i_k$ be a pathway. We create a new graph $G'=(V',E')$, where $V'=V\cup \{i_1',\dots,i_k'\}$ (here, $i_j'$ stands for the copy of vertex $i_j$), and $E'$ expresses as $E'=(E\setminus \{(i_0,i_1)\}) \cup C_1 \cup C_2$, where
$$
C_1=\{(i_0,i_1'\} \cup \{(i_j',i_{j+1}'),\ j=1\dots k-1\},
$$
corresponds to the path and
$$
C_2=\bigcup_{j=1}^{k-1} \cup \{(i_j',\ell),\ (i_j,\ell)\in E\hbox{ and }\ell\neq i_{j+1}\} \cup
\{(i_k',\ell),\ (i_k,\ell)\in E\},
$$
corresponds to the edges needed in order to keep the same sets of paths.

Now, for counting the number of times a given pathway $C$ is taken, it is sufficient to count the number of times transition $(i'_{k-1}, i'_k)$ is taken. 

Consider the following example in which we want to count pathway $C=1\rightarrow 2\rightarrow 3$.

\begin{minipage}{7cm}
\centerline{Original graph $G$}
\includegraphics[width=7cm]{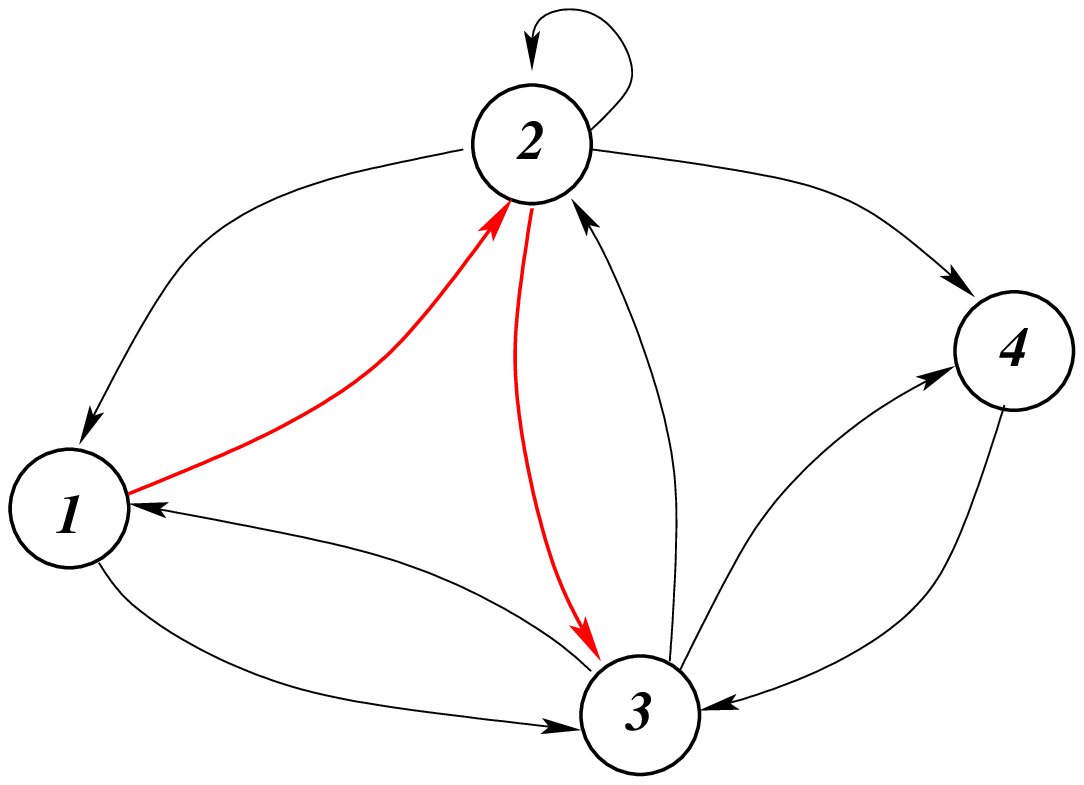}
\end{minipage}
\begin{minipage}{7cm}
\centerline{Modified graph $G'$}
\includegraphics[width=7cm]{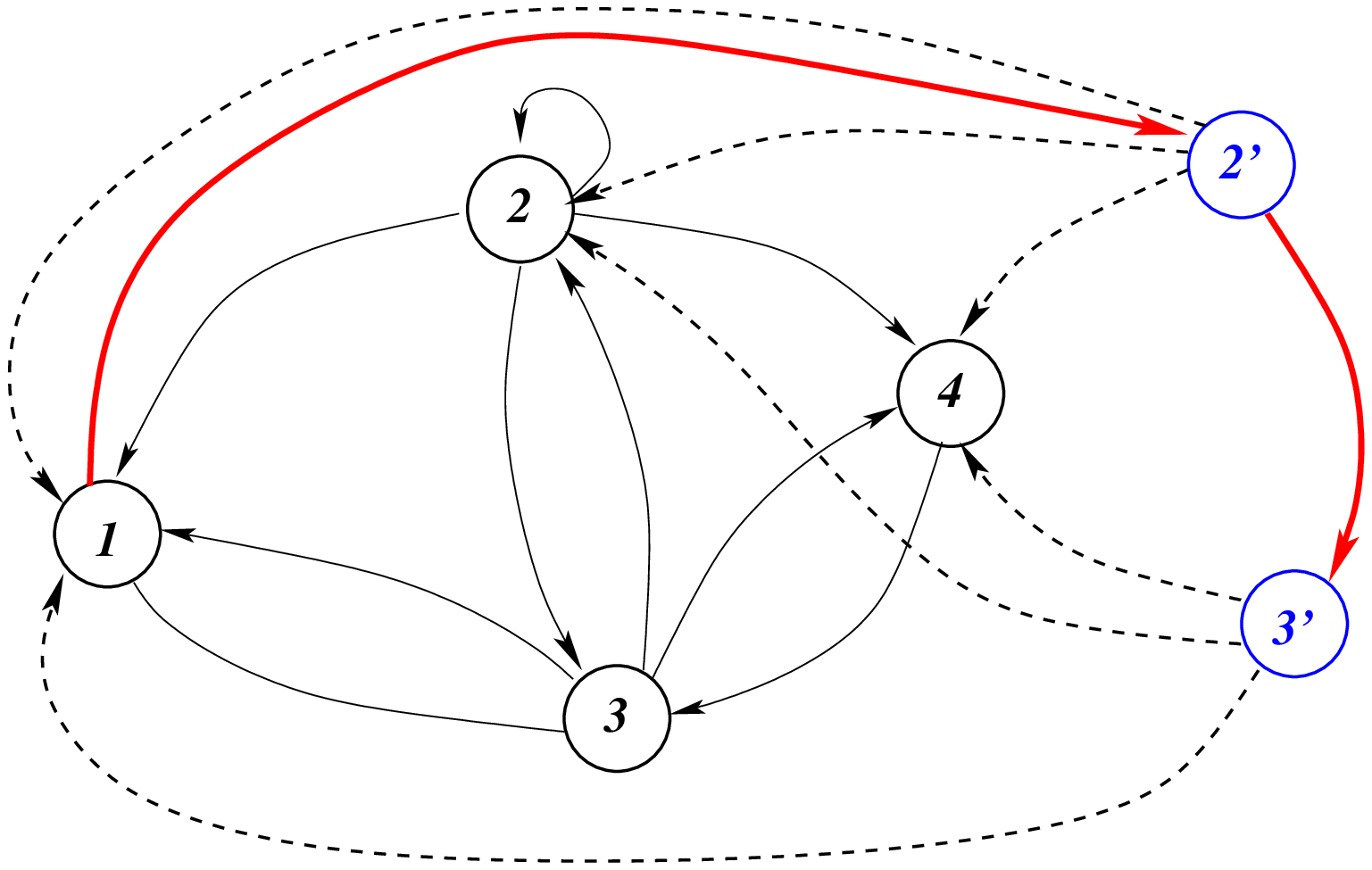}
\end{minipage}

\section{Numerical experiments}
    Here we give two different ways of using the results of this paper. 
    
    The first use is when all the parameter of the model (namely its underlying graph and the probabilistic model of each transition) are fully determined. The model is then used as a simulator and provides some theoretical evidences that may be confirmed later (typically by biological experiments for biological networks).
    
    The second application is a help in the modelling process. Here, the graph is known entirely but the probabilistic model is just partially known. By adding some experimental observed knowledge (an evaluation of a possible measure), it is possible to get some information on the unknown.

We provide some results when the probabilistic model is the simplest one, i.e., the probability of taking one edge is fixed. We will consider a matrix of real numbers in the place of a matrix of operators.

\subsection{Simulation}
Consider the following graph, transition matrix operator $\mathbb{T}$ and cost matrix $C$ (that marks the edge $(1,2)$

\begin{minipage}{5cm}
$$
\includegraphics[width=2.7cm]{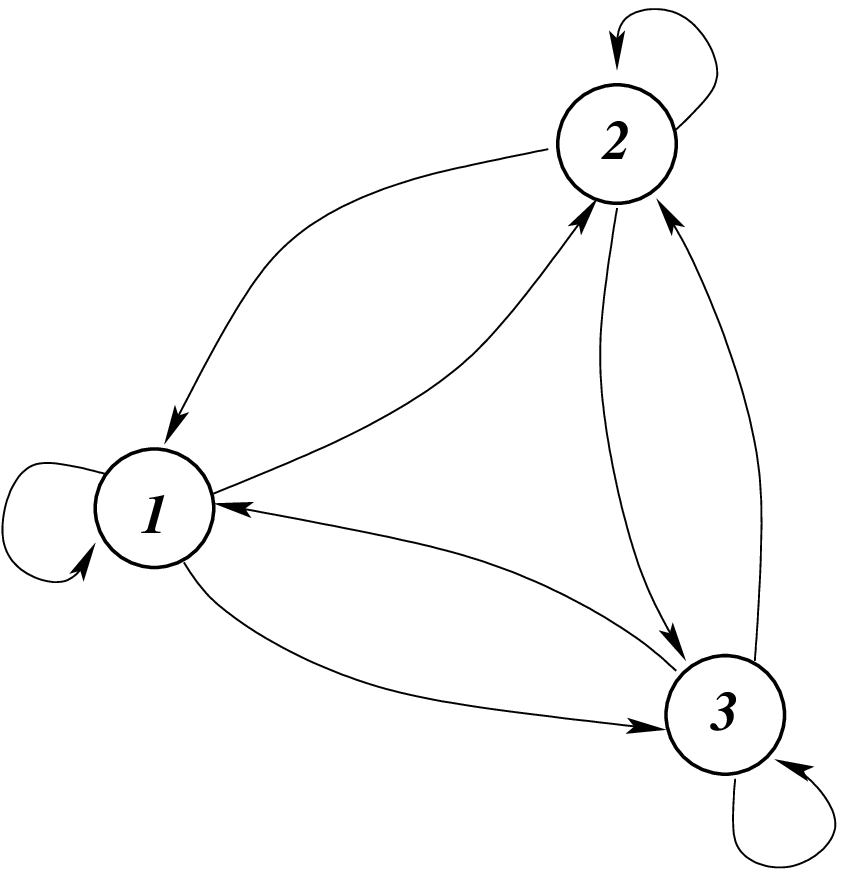}
$$
\end{minipage}
\begin{minipage}{10cm}
$$
\mathbb{T}=\left ( \matrix{
0.4 & 0.2 & 0.4\cr
0.7 & 0.2 & 0.1\cr
0.2 & 0.7 & 0.1\cr
} \right )\qquad
C=\left ( \matrix{
0 & 1 & 0\cr
0 & 0 & 0\cr
0 & 0 & 0\cr
} \right ).
$$
\end{minipage}

The dominant eigenvalue $\lambda(u)$ of the marked matrix operator $\mathbb{T}(u)$ equals
$$
\lambda(u)=\frac{K(u)^{2/3}+208+168u+14K(u)^{1/3}}{60K(u)^{1/3}}, 
$$
where $K(u)=20276+2448u+12\sqrt{2792481 + 537960 u - 80688 u^2  - 32928u^3}$.
Thus $\lambda'(1)=0.0896$ which indicates that the edge $(1,2)$ is taken approximately $8.9\%$ of times in long paths in the graph.

\subsection{Modelling}
Consider now the following graph, with a transition matrix operator $\mathbb{T}$ partially known

\begin{minipage}{5cm}
$$
\includegraphics[width=2.7cm]{automatebase}
$$
\end{minipage}
\begin{minipage}{10cm}
$$
\mathbb{T}=\left ( \matrix{
0 & x & 1-x\cr
y & 0 & 1-y\cr
z & 1-z & 0\cr
} \right ).
$$
\end{minipage}

Nevertheless, it is observed that edge $(1,3)$ is taken in average between $30\%$ and $40\%$ of times while edge $(3,1)$ is taken in average between $20\%$ and $30\%$ of times. The question is ``does it give any information for $x$, $y$ and $z$ ?''. Obviously, yes. This problem can be seen as a constraint problem with 3 variables over the continuous domains $D_x=D_y=D_z=]0,1[$ with some constraints that are inequalities based on the observations. Eigenvalues $\lambda_{(1,3)}(u)$ and $\lambda_{(3,1)}(u)$ associated to the observations express symbolically with $x$, $y$ and $z$. One has the constraints
$$
0.3\leq \lambda'_{(1,3)}(1) \leq 0.4,\qquad
0.2\leq \lambda'_{(3,1)}(1) \leq 0.3.
$$
We discretize the three domains and keep only 100 regularly spaced values and then operate an exhaustive determination of all possible graphs (among the $100^3$ possible ones) that satisfies the first constraint, the second constraint and both the two constraints. This is represented by the following 3D pictures.

\vskip 1cm
\begin{minipage}{5cm}
\kern -1cm\includegraphics[width=6cm]{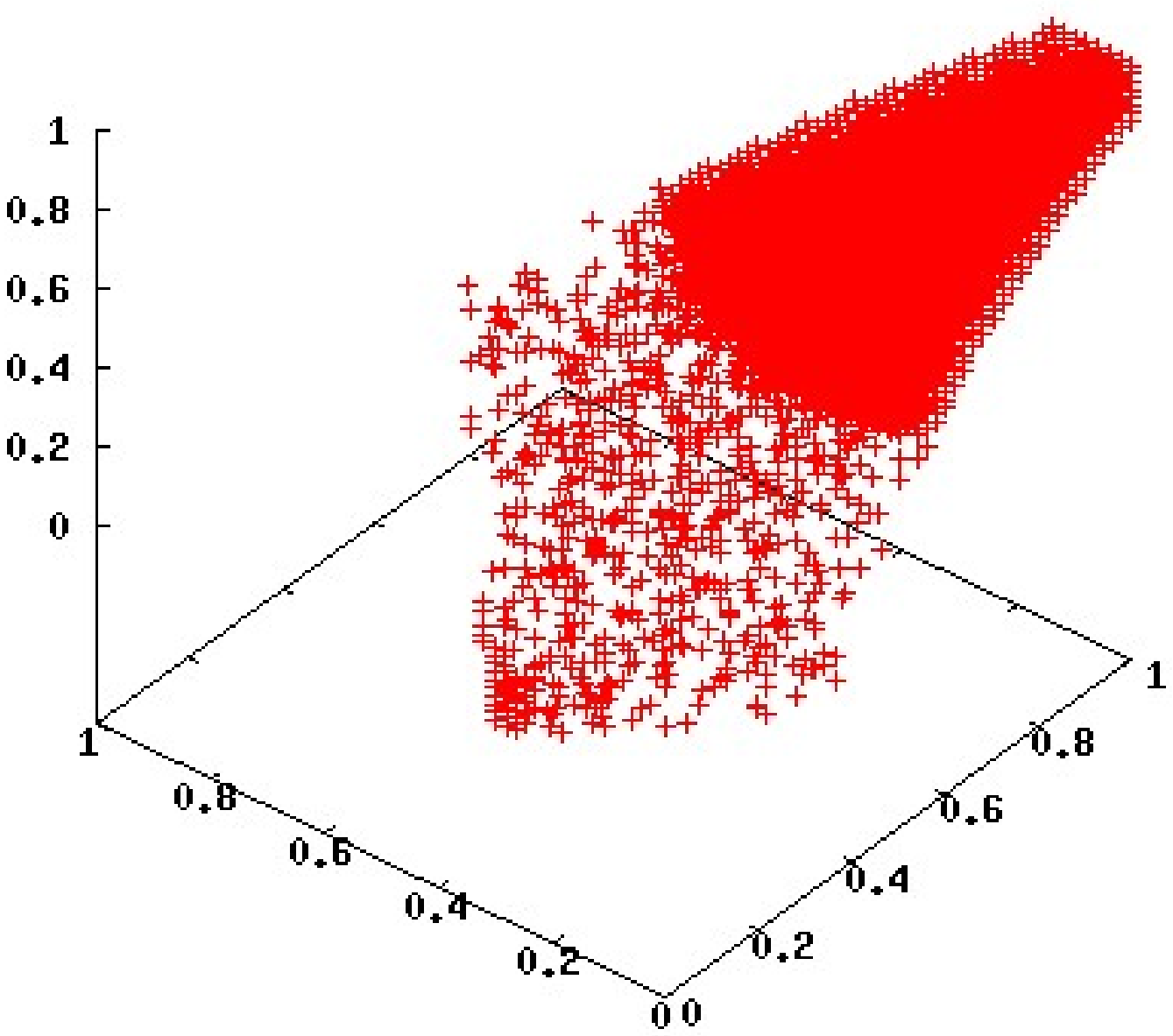}
\centerline{First constraint}
\end{minipage}
\begin{minipage}{5cm}
\kern -1cm\includegraphics[width=6cm]{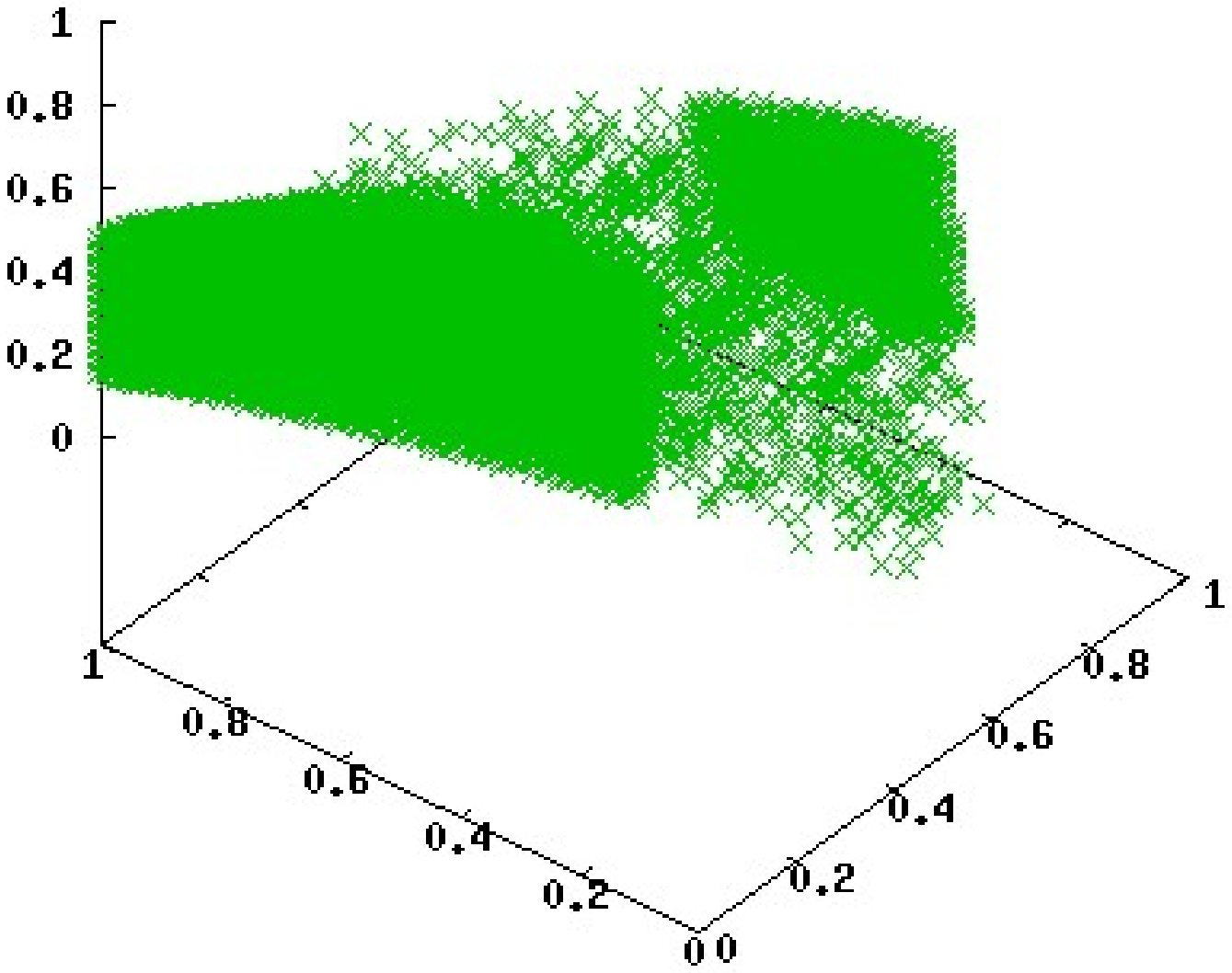}
\centerline{Second constraint}
\end{minipage}
\begin{minipage}{5cm}
\kern -1cm\includegraphics[width=6cm]{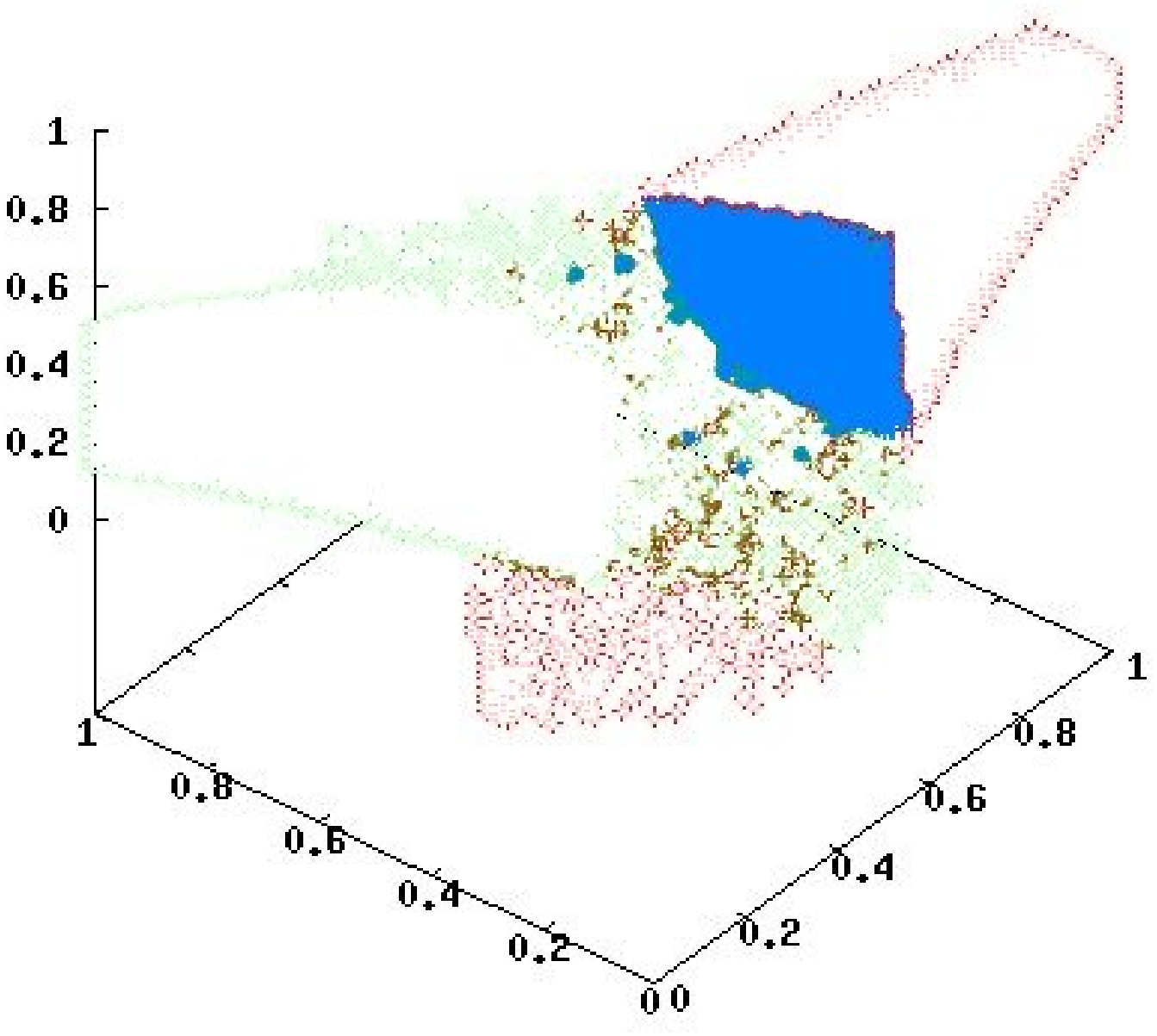}
\centerline{Both}
\end{minipage}

The graphs of interest correspond to the common part, a small part of the entire cube $[0,1]^3$.

\section{Conclusion and perspectives}

Our study illustrates toll based measure schemes in a large class of probabilistic model of graph. It shows (i) wide generality on both measure and probabilistic model and (ii) entails theoretical results on graphs that might be useful in various applications with a special emphasis on biological networks application. Our quantitative analysis characterizes the major impact of other measure scheme, such those related to the waiting time before taking a given transition. Note here that they surely do not obey asymptotically to a Gaussian law.
Behind these theoretical items, our model rises various perspectives. One of them considers a multivariate framework where the study concerns the asymptotic behavior of a vector of different measures. In this context, and for a vector of toll based measure, the work of Bender, \textit{et al.}~\cite{BeRiWi83,bk93} might be particularly useful. Finally, for a modeling purpose, the determination of $n$ unknown parameters of the probabilistic model consist in finding an appropriate part of the hypercube $[0,1]^n$. Here shows the importance to design efficient methods that find good approximations of this part. The local search community develops several methods such as tabu search or genetic algorithms that are still to be considered for investigating graph properties.

\end{document}